# Imaging localized plasmon resonances in vacancy doped $Cu_{3-x}P$ semiconductor nanocrystals with STEM-EELS

Giovanni Bertoni,[1,2] Rosaria Brescia,[3] Luca De Trizio,[2] Liberato Manna,[2] Quentin Ramasse[4]

[1] CNR - Istituto Nanoscienze, Via Campi 213/A, 41125 Modena, Italy

[2] Dept. Nanochemistry, Istituto Italiano di Tecnologia, Via Morego 30, 16163 Genova, Italy

[3] Electron Microscopy Facility, Istituto Italiano di Tecnologia, Via Morego 30, 16163 Genova, Italy

[4] SuperSTEM, SciTech Daresbury Science and Innovation Campus, Keckwick Lane, Daresbury WA4 4AD, UK

**Abstract**

Copper binary compounds are often intrinsic p-type semiconductors due to the presence of Cu(I) vacancies, with corresponding hole carriers in the valence band. If the free carrier concentration is high enough, localized plasmon resonances can be sustained in nanocrystals, with frequencies in the infra-red (<1 eV), with respect to the typical resonances seen in the visible range in the case of metals (Ag, Au, …). The localization of the resonances can be demonstrated with scanning transmission electron energy loss spectroscopy (STEM-EELS) by combining high spatial and high energy resolutions. Here we demonstrate that Cu(I) vacancies can be directly measured from the STEM images in $Cu_{3-x}P$ hexagonal nanocrystals. Two localized resonances can be seen from STEM-EELS, which are in agreement with the resonances calculated from the vacancy concentration obtained from the STEM.

**Introduction**

In the last decade, electron energy loss spectroscopy in transmission scanning spectroscopy (STEM-EELS) has permitted to directly visualize localized plasmon resonances (or LPRs) in metallic nanostructures (such as Ag and Au nanostructures).[1,2,3] This has become possible by combing the high spatial resolution and high beam current from probe aberration correctors with the high resolution in the spectrum obtained with monochromators (such as the Wien filter). These measurements became more and more routinely available, and a verification of the predictions from theory (such as Drude theory, Mie theory, or boundary element methods, …)[4] more accurate.

Beyond metallic nanoparticles, doped semiconducting nanostructures may present plasmonic features deriving from the collective excitations of free carriers. Typically, these are seen as peaks in the optical absorption spectrum. Copper binary compounds are suitable candidates due the possibility to tune the copper vacancies to create hole carriers in the valence band, which may sustain the LPRs. Examples of similar systems are $Cu_{2-x}S$, or $Cu_{2-x}Se$ colloidal quantum dots,[5,6,7] in which the increasing number of Cu vacancies results in higher hole concentrations, and consequently in a shift of the LPRs at higher frequencies, as measured from light absorption. However, these measurements are limited to an average information of the sample, with no information of the spatial localization of the resonance in a single crystal. Recently, also $Cu_{3-x}P$ nanocrystals were proposed as plasmonic semiconductors for using in photovoltaics or photodetectors, due the presence of Cu vacancies and consequently hole carriers in the valence band, with a broad plasmon peak in optical absorption at around 1500 nm.[8,9,10]

However, two aspects make the visualization of localized resonances in semiconductors a challenging task in STEM-EELS: a) the lower cross-section of the electronic coupling with respect to the photon coupling, b) the lower carrier density, shifting the resonances to lower energies (usually in the infra-red, where the tail of the elastic peak dominates) and lowering the probability of excitation (if the carrier concentration is too low, the resonances cannot be sustained). Several tries were done on Cu based semiconducting nanocrystals,[11,12] but the results were limited to average spectra, due to low signal to noise ratio, with no direct evidence of the localization of the resonances, which could be derived by taking an eels spectrum at each point of a scan across the particle (called spectrum image, or SI).

In this manuscript we demonstrate that two LPRs are visible in STEM-EELS in a $Cu_{3-x}P$ hexagonal nanoplatelets (diameter 50 nm, height 10 nm), one at about 0.6 eV and one at about 0.8 eV. These are due to the generated hole carriers due to Cu vacancies in the hexagonal P63cm structure. The Cu vacancies are verified directly from the ADF image, resulting in a very good match of the simulated image and the experiment. The two

LPRs matches with the two dominant modes from a boundary element simulation of the EELS probability based on a Drude model for the dielectric function.

**Experimental**

Atomic resolution images of the $Cu_{3-x}P$ nanocrystals were acquired in annular dark field (ADF) STEM on a Nion UltraSTEM100 microscope, equipped with a probe aberration corrector (SuperSTEM, Daresbury, UK) and an Enfinium high resolution spectrometer (Gatan, Inc.). The convergence semi-angle was 31 mrad and the inner cutoff angle of the STEM detector 95 mrad. For better signal to noise ratio and reduced drift distortions in the ADF image, several fast scans (>20) were acquired and summed after cross correlation correction. For the simulation of the ADF image, we used the orthorhombic notation of the pristine hexagonal cell (**a'** = **a**, **b'** = 2**b** – **a**, and **c'** = **c**). The hexagonal cell (**a**, **b**, **c**) was taken from ICSD#15056 (P63cm, s.g. 185) considering a = b = 0.696 nm, and c = 0,714 nm.[13] The ADF image was calculated by using the software DrProbe.[14] We took into account the finite size of the probe by propagating a Gaussian source of 100 pm width (FWHM). Electron energy loss spectra, in the form of 3D (*x, y, E*) datasets (i.e. spectrum image or SI) were acquired at 60 keV on the same machine by exciting the monochromator (Wien Filter), resulting in a FWHM of the elastic peak or zero-loss peak (ZL) of 0.024 eV. The acquisition at 60 keV reduces the retardation effects on the energy loss spectrum at low energy.[15,16] Moreover, the SI acquisition reduces the damage of the sample by spreading the current across a large sample area. The spectra were acquired in the Dual EELS mode, permitting the acquisition of two simultaneous spectra on the CCD camera of the spectrometer. One spectrum (low-loss) contains the ZL peak acquired at short exposure (10 msec), while the other spectrum (high-loss) contains the region of the band-gap acquired at longer exposure (100 msec). The total acquisition time for the two spectra at one pixel was about 110 msec. The energy loss dispersion was fixed to 0.002 eV per channel. The low-loss spectrum is used to carefully align the spectra at sub-pixel level in the energy scale by fitting a Gaussian peak under the ZL peak. The result is applied to the corresponding high-loss spectra. We assume no drift in the energy loss during the fast drift tube change of the spectrometer from low-loss to high-loss region in the acquisition of a single pixel of the SI.

The simulation of the EEL spectra and energy loss maps were performed with the MNPBEM toolbox,[17] using the retarded approximation. A hexagonal platelet with rounded edges was used to approximate the shape of the nanocrystals, and a thickness of 10 nm was considered. A 3 nm thin square plate made of amorphous carbon with a constant dielectric function (n = 2.45) was placed below the hexagonal platelet to approximate the presence of the support carbon film of the TEM grid.

**Results and discussion**

**STEM imaging**

The high-resolution image from a [001] projection of a single $Cu_{3-x}P$ nanoplatelet is presented figure 1a-b. Clearly, a hexagonal structure compatible with $Cu_{3-x}P$ is visible. The atomic structure is presented in Figure 1c, in which the inequivalent Cu positions are indicated with different colors. The simulated ADF image corresponding to 2x2 orthorhombic cells obtained from a full $Cu_3P$ stoichiometry with all Cu occupations set to 1.0 is presented in Figure 1d. As can be seen, a full stoichiometric $Cu_3P$ does not match the contrast in the experiment. Indeed, we have previously calculated by density function theory (DFT) the vacancy formation energy ($E_V$) for the four Cu sites. The corresponding $E_V$ values are negative for Cu1 and Cu2 sites, and positive for Cu3 and Cu4 sites.[9] According to these results, we set the occupation of Cu1 and Cu2 sites to 0.9, while keeping the occupation of Cu3 and Cu4 at 1.0. This corresponds to a stoichiometry $Cu_{2.79}P$ which gives a very good match between the ADF simulation and the experiment (Figure 1f). This means 2.4 Cu vacancies in the orthorhombic cell (full line in the figure), or 1.2 in the hexagonal unit cell (dashed line). This confirms the prediction of Cu vacancies in this system from DFT calculations previously reported.[9,18] It further proves the capability of ADF as a quantitative technique, if all the relevant parameters of the acquisition are well known.

**Hole carriers and localized plasmons**

The second step is to verify if the Cu vacancies generates enough hole carriers to sustain localized plasmon modes. By considering every vacancy is generating a hole carrier ($n_h = n_V$), we expect $n_h \sim 4.0 \cdot 10^{21}$ cm$^{-3}$. In the case of a crystal with d = 50 nm and h = 10 nm, this means around 80000 hole carriers in a single nanocrystal, enough to sustain a LPR.[5] However, we expect the resonances at low energy, *i.e.* in the infra-red

(around 0.8 eV) according to the optical measurements.[9] For these reasons we opted for dual EELS acquisition, for a careful measurement of the position of the ZL peak and a good statistic in the expected plasmonic region. To avoid to damage the beam sensitive crystals, we opted to scan along a smaller area of the crystal, starting from the center of the hexagon and crossing the edge. The results are presented in Figure 2. Figure 2a show two integrated spectra at the center and at the edge of the nanocrystal. Two peaks are clearly visible at approximately 0.6 eV and 0.8 eV, corresponding to two different LPRs. Indeed, by subtracting a power-law background approximating the tail of the ZL peak, two intensity maps corresponding to the two LPRs can be extracted (figure 2b). Clearly, the LPR at 0.6 eV is located at the edge of the hexagon, while the LPR at 0.8 eV is located at the center.

To further confirm the plasmonic nature of the two peaks, we performed a simulation of the EEL spectra and spectral maps using the boundary element method (MNPBEM Toolbox).[17] Unfortunately, an experimental dielectric constant $\varepsilon(E)$ for $Cu_{3-x}P$ is not available from literature. For this reason, we built a Drude model, by calculating the plasmon (bulk) frequency from the estimated carrier density $n_h$, according to the formula:

$$\omega_p = \sqrt{\frac{n_h e^2}{\varepsilon_0 m_h}}$$

with $\varepsilon_0$ the vacuum permeability, $e$ the electron charge, and $m_h$ the hole mass ($m_h$ = 1.4 $m_0$).[8] the complex dielectric function $\varepsilon(E)$ was calculated as:

$$\varepsilon(E) = 1 - \frac{\hbar^2 \omega_p^2}{E(E + i\gamma)}$$

The derived dielectric function was used as input in the retarded simulation ($\gamma$ = 0.02 eV). The results of the EELS simulation are shown in Figure 3. Figure 3a shows the loss probability measured at three different positions: at one corner of the hexagon, at the center, and at one edge. Figure 3b shows the corresponding intensity maps from the two peaks in the loss probability. Clearly, there are two intense modes that correspond to the two peaks observed experimentally. The first one at 0.53 eV is localized at the edges and corners of the hexagon, while the one at 0.84 eV is located at the center. To take into account the presence of the support film, the hexagon in the simulation was laid on a 3 nm thin amorphous substrate (the nominal thickness of the ultrathin carbon grids of the experiment), and approximated with a square plate of 150 nm edge width. Retardation effects were included in the simulation, to account for the red shift of the peaks due to the substrate. Please note that within a Drude model approximation, the simulated LPRs are probably more intense than the real ones. The simulation is to be intended for a qualitative comparison. A more realistic model of the dielectric function is needed to accurately match intensities and energy positions of the LPR peaks. Nevertheless, the calculation is in very good agreement with the experimental findings, proving indeed that hole carriers can sustain localized plasmonic resonances in this system.

**Conclusions**

We have verified the presence of Cu vacancies in single semiconducting $Cu_{3-x}P$ nanocrystals. From the ADF image we demonstrated that the Cu vacancies are located preferentially at the Cu1 and Cu2 sites as predicted from theory, resulting in a $Cu_{2.79}P$ composition. Moreover, the presence of Cu vacancies generates free hole carriers which can be collectively excited, resulting in localized plasmon resonances (LPRs) in the infra-red region, as seen in the EELS maps acquired across the edge of a $Cu_{3-x}P$ hexagonal nanocrystal. The hexagonal $Cu_{3-x}P$ nanocrystals show two principal resonances at the edge and at the center as confirmed by a simple Drude model for the dielectric function.

**Figures**

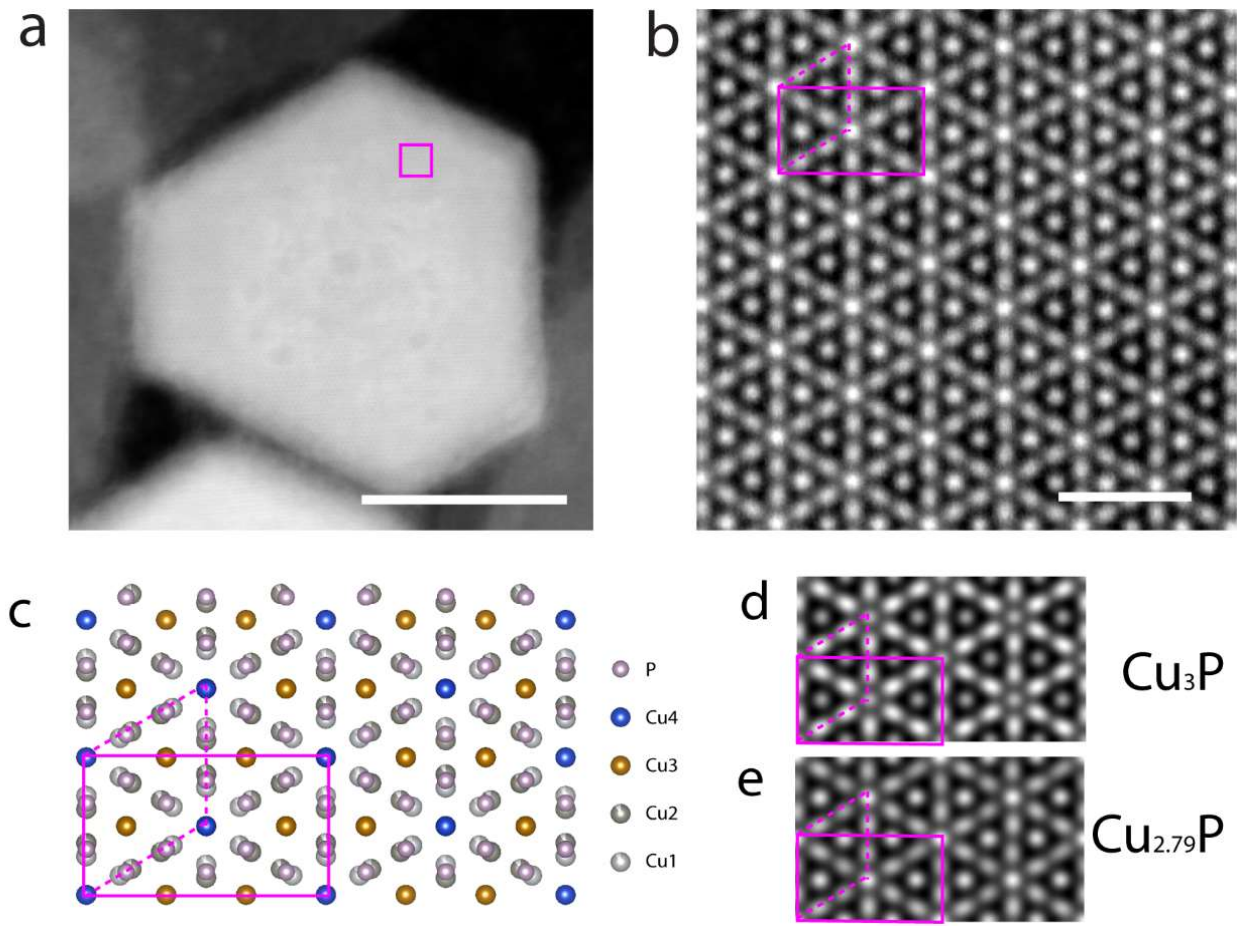

**Figure 1.** a) ADF image of a single $Cu_{3-x}P$ nanocrystal. Scale bar 20 nm. b) high magnification image of the region indicated in a). The hexagonal P63cm unit cell (ICSD#15056) (purple dashed line) and the orthorhombic cell (purple solid line) used for simulation are indicated. Scale bar 1 nm. c) Plot of the P63cm structure with color codes for the inequivalent atoms. A 0.9 occupation in Cu1 and Cu2 is indicated, while Cu3 and Cu4 have occupation 1.0. d) Simulated ADF image for the full stoichiometry $Cu_3P$ (no vacancies). e) Simulated ADF image for the structure with partial occupation at Cu1 and Cu2 sites, resulting in $Cu_{2.79}P$ stoichiometry.

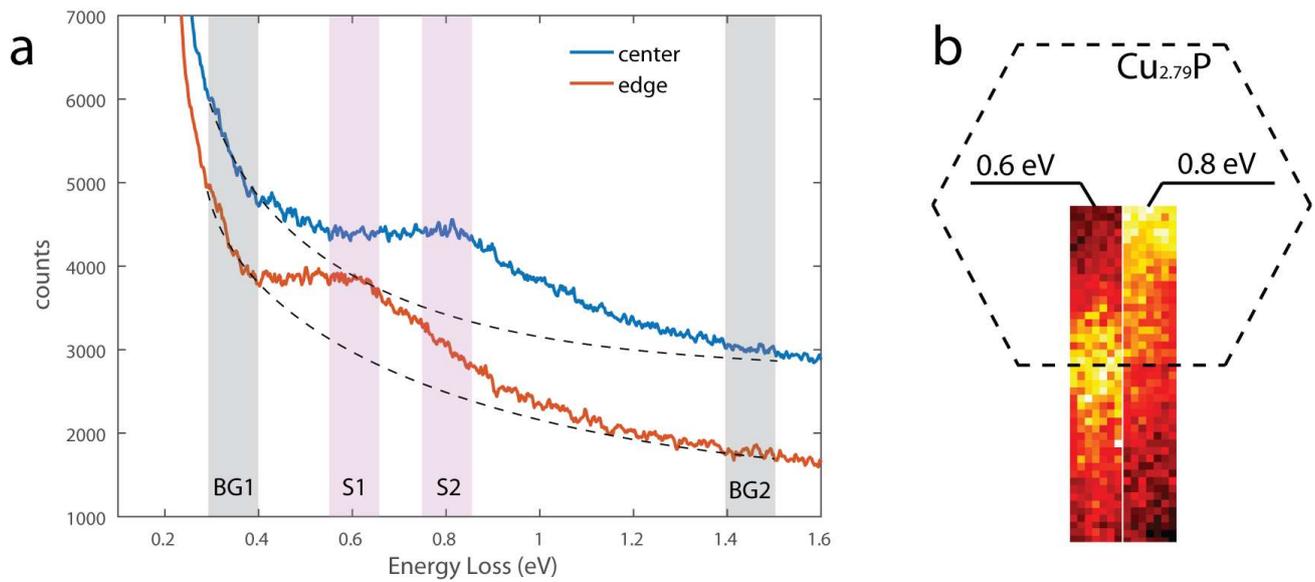

**Figure 2.** a) Representative EEL spectra from a $Cu_{3-x}P$ nanocrystal. Two main resonances can be seen localized at the edge (orange line) and at the center of the crystal (blue line). b) Intensity maps from the two resonances, obtained after background subtraction of a power-law function, extrapolated from the two fitting regions BG1 and BG2. The intensity maps refer to the integrated regions S1 and S2, respectively. All the background and signal regions have 0.1 eV width.

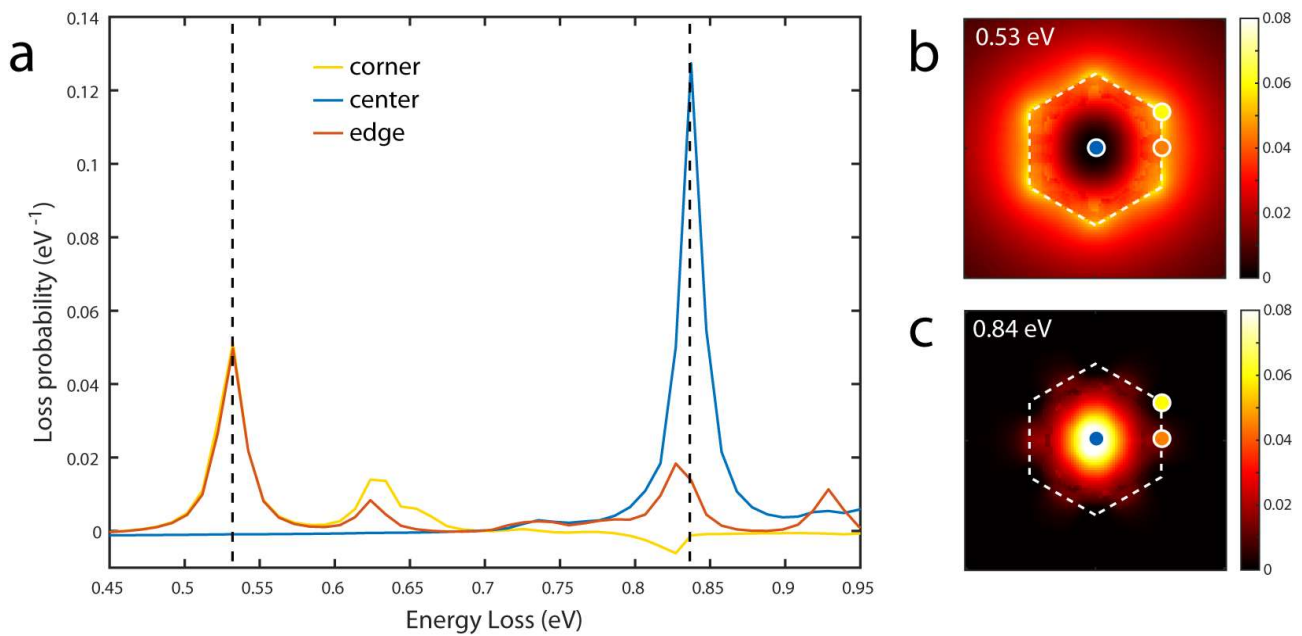

**Figure 3.** MNPBEM simulated EEL spectra from the Drude model of the dielectric function in eq.2. a) EEL spectra simulated at a corner, at the center, and at the edge of a $Cu_{2.79}P$ nanocrystal. b-c) Corresponding

intensity maps at the energy losses of the two peaks indicated in a) with dashed lines. The color circles correspond to the spatial positions for the spectra in a). The shape of the crystal is sketched with a dashed white line.